%% file: main.tex
\documentclass[natbib,vecphys]{svmult}
\usepackage{enumerate}
\usepackage{graphicx}     
\usepackage{amsmath}

\usepackage{subfigure}
\usepackage{amsfonts}
\newcommand{\beq}{\begin{equation}}
\newcommand{\eeq}{\end{equation}}
\newcommand{\be}{\begin{eqnarray}}
\newcommand{\bes}{\begin{eqnarray*}}
\newcommand{\ee}{\end{eqnarray}}
\newcommand{\ees}{\end{eqnarray*}}

\newcommand{\RR}{\mathbb{R}}

\begin{document}

\title*{Representing complex data using localized principal components with application to astronomical data}
\titlerunning{Representing complex data using localized principal components}
\author{Jochen Einbeck\inst{1}\and Ludger Evers\inst{2}\and Coryn Bailer-Jones\inst{3}}
\authorrunning{Einbeck et al.}
\institute{Department of Mathematical Sciences, Durham University, Science Laboratories, South Road, Durham DH1 3LE, United Kingdom, \texttt{jochen.einbeck@durham.ac.uk}\and
Department of Mathematics, University of Bristol, University Walk, BS8 1TW, United Kingdom, \texttt{l.evers@bris.ac.uk} \and
Max-Planck-Institut f\"ur Astronomie, K\"onigstuhl 17, 69117 Heidelberg, Germany, \texttt{calj@mpia-hd.mpg.de}}

\maketitle

\
\begin{center}
in {\em Principal Manifolds for Data Visualization and Dimension Reduction},\\
A. Gorban, B. Kegl, D. Wunsch, and A. Zinovyev (eds),\\
Lecture Notes in Computational Science and Engineering,\\
Springer, 2007, pp.~180--204
\end{center}

\begin{abstract}

Often the relation between the variables constituting a multivariate data 
space might be characterized by one or more of the  terms: ``nonlinear'', 
``branched'', ``disconnected'', ``bended'', ``curved'', ``heterogeneous'', or, more general, ``complex''. In these cases, simple principal component analysis (PCA) as a tool for dimension reduction can fail badly. Of the many alternative 
approaches proposed so far, local approximations of PCA are among the most 
promising. This paper will give a short review of localized versions of PCA, 
focusing on local principal curves and local partitioning algorithms. Furthermore we discuss projections other than the local principal components. When performing local dimension reduction for regression or classification problems it is important to focus not only on the manifold structure of the covariates, but also on the response variable(s). Local principal components only achieve the former, whereas localized regression approaches concentrate on the latter. Local projection directions derived from the partial least squares (PLS) algorithm offer an interesting trade-off  between these two objectives.  

We apply these methods to several real data sets. In particular, we consider simulated astrophysical data from the future  Galactic survey mission Gaia.
\end{abstract}

\newpage

\input{sec1}

\input{sec2}

\input{sec4}
\input{sec5}

\input{sec6}

% I'm not 100% sure whether Springer lets us use Bibtex
\bibliographystyle{plain} % ???????
\bibliography{pc_gaia}

%\printindex
\end{document}

%% file: sec1.tex
\section{Introduction}

Principal component analysis is a well established tool for dimension reduction. For data $\vec{X} = (\vec{x}_1', \ldots, \vec{x}_n')'$ with $\vec{x}_i\in\mathbb{R}^p$ the principal components provide a sequence of best linear approximations to it. Let $\vec{\Sigma}$ be the empirical covariance matrix of $\vec{X}$, then  the principal components are given by the eigen decomposition 
\be
\label{pcdecom}
\vec{\Sigma} = \vec{\Gamma\Lambda\Gamma}'
\ee
where $\vec{\Lambda} = \mbox{diag}(\lambda_1, \ldots, \lambda_p)$ is a diagonal matrix containing the ordered eigenvalues of $\vec{\Sigma}$, with $\lambda_1 \ge \ldots \ge \lambda_p$,  and  $\vec{\Gamma}$ is an orthogonal matrix. The columns of $\vec{\Gamma} =(\vec{\gamma}_1, \ldots, \vec{\gamma}_p)$ are the eigenvectors of $\vec{\Sigma}$ and are called \textsl{principal component loadings}. The first loading $\vec{\gamma}_1$ maximizes the variance of the \textsl{scores} $\vec{X}\vec{\gamma}$ over all $\vec{\gamma} \in \mathbb{R}^p$ with $\|\vec{\gamma}\|=1$, the second loading $\vec{\gamma}_2$ maximizes the variance of $\vec{X}\vec{\gamma}$ over all $\vec{\gamma} \in \mathbb{R}^p$ with $\|\vec{\gamma}\|=1$ which are orthogonal to $\vec{\gamma}_1$, and so on. 

\begin{figure}
\hspace*{-0.5cm}
\subfigure[10th December 2006]{\includegraphics[width=6.4cm,angle=270]{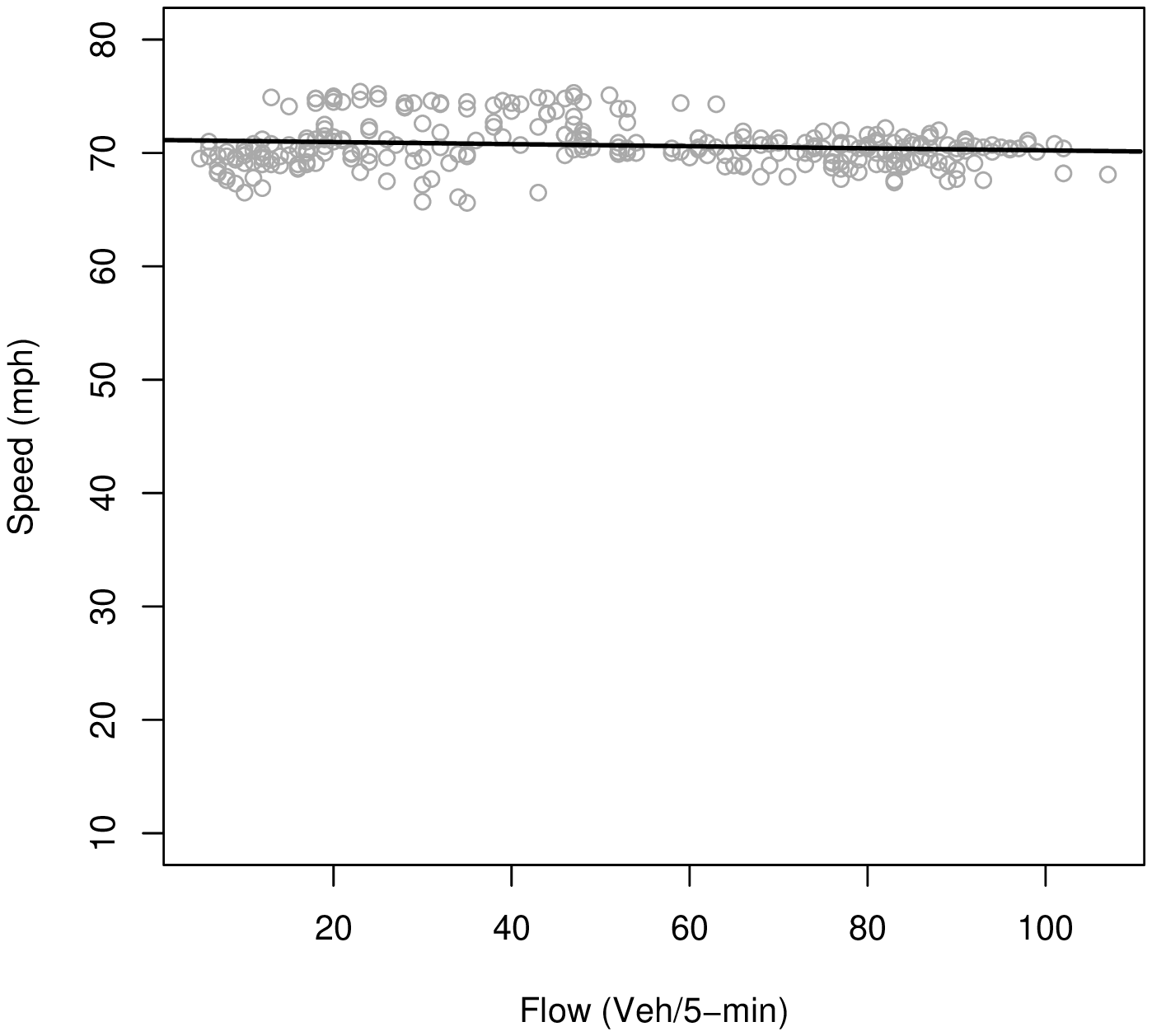}}
\hspace*{-0.4cm}
\subfigure[7th December 2006]{\includegraphics[width=6.4cm,angle=270]{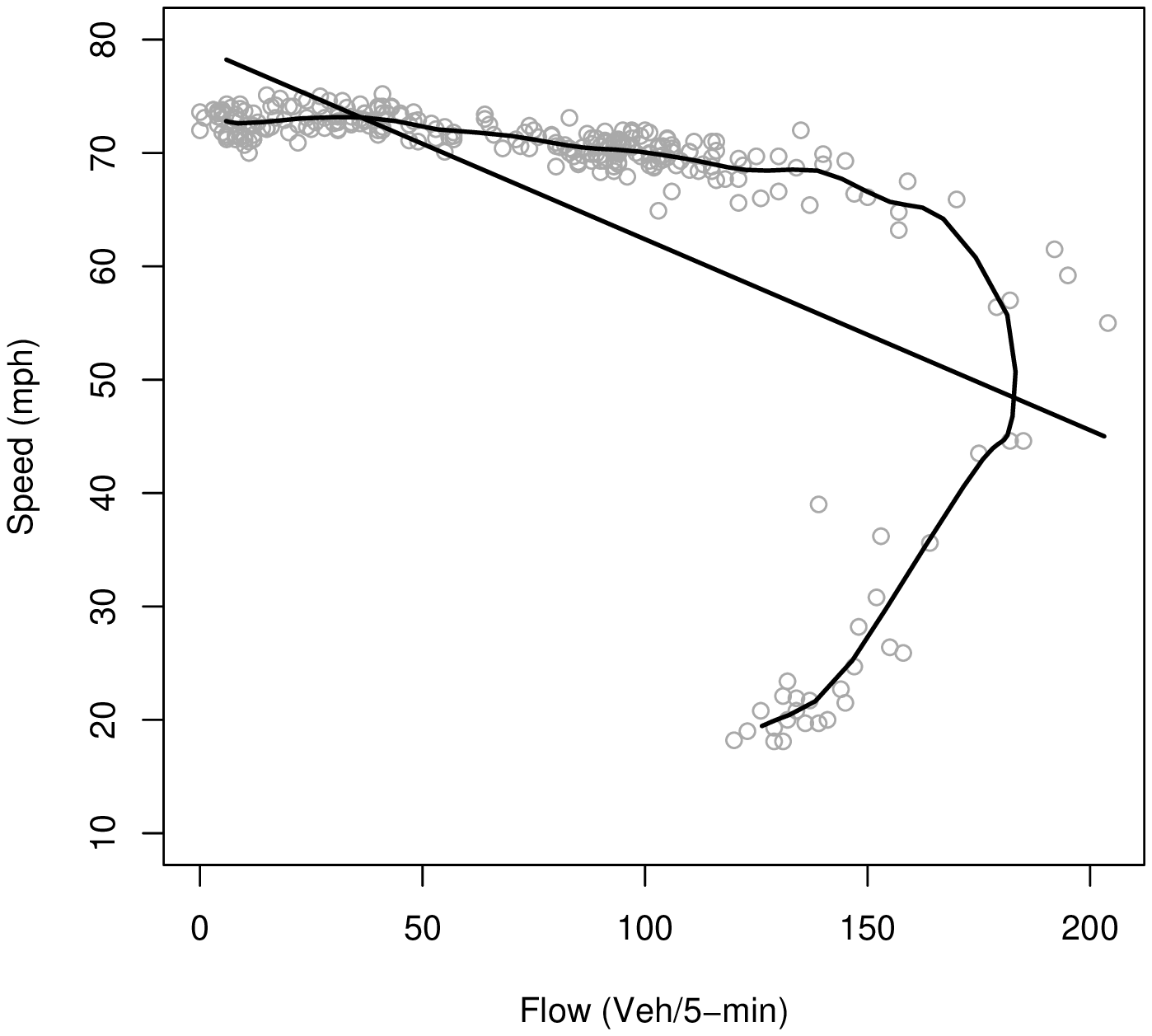}}
         \caption{\label{dez710} Speed flow diagrams for Freeway 4-W on 10th (a) and 7th (b) of December 2006, with principal component lines (straight lines) and a principal curve through the latter data set (smooth curve).}
\end{figure}

As an example, we consider a speed-flow diagram recorded on the freeway 4-W in Contra Costa County,  California (figure \ref{dez710}(a)).\footnote{The  data are taken from the database PemS \cite{PemS:04}.} The displayed points correspond to the average speed and flow over 5 minute intervals. The data were collected on 10th December 2006; this was a Sunday, so the traffic flow was quite low, not exceeding 1300 vehicles per hour.  Most cars went at a speed close to the speed limit. One can easily imagine a first principal component line $\vec{g}(\eta)=\bar{\vec{x}} +\eta\cdot\vec{\gamma}_1$ (with overall mean $\bar{\vec{x}}$)  fitted through the 
data cloud, capturing the main part  (here: $99.54\%$) of the variance in this data set (figure \ref{dez710}(a)). Clearly, the projection indices $\eta_i=(\vec{x}_i-\bar{\vec{x}})^T\vec{\gamma}_1$ of the data projected onto this line are good indicators for the positions of the data points within the data cloud.

However, the application of principal component analysis postulates implicitly some form of linearity. More precisely, one assumes that the data cloud is directed, and that the data points can be well approximated by their projections onto the line $\vec{g}$ corresponding to the first principal component (or in general the affine hyperplane corresponding to the first $d$ principal components). This implicit assumption of linear PCA can already be violated  in very simple situations. As an example, consider the speed-flow diagram recorded three days earlier with the same detector at the same location. This was a Thursday and traffic was busy, leading to occasional congestion. As a consequence, cars often had to reduce their speed at higher flows, resulting in the data cloud shown in figure \ref{dez710}(b). This kind of pattern has frequently been reported in the transportation science literature, see e.g. \cite{HalHurBan:92}.

These data are somewhat half-moon shaped and it does not seem to make much sense to speak of some principal direction for this data. The projection indices onto any straight line like the first principal component (figure \ref{dez710}(b)) would be uninformative with respect to the position of the data within this bent data cloud.  This will certainly get worse if the the data cloud is strongly twisted,  if it has crossings,  if it consists of several branches, if the curvature permanently changes, or if there are several disconnected clouds, and so on.

Solutions to problems of this kind are readily available, and fall into two major categories: {\it Nonlinear principal component analysis} (hereafter: NLPCA) and {\it principal curves} (or their multivariate extension, {\it principal manifolds}).  The dividing lines between these two approaches are often rather fuzzy, but the following vague rule can be used to distinguish between them. Principal curves is a nonparametric extension of linear PCA, while NLPCA is a parametric  (or semi-parametric), but nonlinear version of PCA. Most, but not all algorithms belonging to either of the categories accomplish their task in two steps:
\begin{description}
 \item[\bf Projection] Define a dimension reducing transformation  $\vec{f}(\vec{x}): \RR^p \rightarrow \RR^d$
 \item[\bf Reconstruction] Find a mapping back to the data space  $\vec{g}(\vec{\eta}): \RR^d \rightarrow \RR^p$
\end{description}

Here, the  vector $\vec{\eta}$ represents a $d-$ dimensional parameterization of the $p-$ dimensional data space. Summarizing, the reconstructed curve is given by $\vec{g}(\vec{f}(\vec{x}))$, which is mostly found by (implicitly or explicitly) minimizing the reconstruction error \begin{equation}
\label{Xmin}
 \mathbb{E} \|\vec{x}-\vec{g}(\vec{f}(\vec{x}))\|^2,
\end{equation}
or its empirical counterpart
\be
\label{EmpXmin}
\sum_{i=1}^n \|\vec{x}_i-\vec{g}(\vec{f}(\vec{x}_i))\|^2.\label{XminEmp}
\ee

PCA can be seen as a special case of the above algorithm. It follows directly from the extremal properties of the eigenvectors that the first $d$ principal components define the \emph{linear}, $d$-dimensional functions minimizing equations (\ref{Xmin}) or (\ref{XminEmp}). In the special case of PCA $\vec{f}(\vec{x})=\left(\vec{\gamma}_1\cdots\vec{\gamma}_d\right)'(\vec{x}-\vec{\mu})$, $\vec{g}(\vec{\eta})=\vec{\mu}+\left(\vec{\gamma}_1\cdots\vec{\gamma}_d\right)\vec{\eta}$, and $(\vec{g}\circ\vec{f})(\vec{x})=\vec{\mu}+\left(\vec{\gamma}_1\cdots\vec{\gamma}_d\right)\left(\vec{\gamma}_1\cdots\vec{\gamma}_d\right)'(\vec{x}-\vec{\mu})$, where $\vec{\mu}=\mathbb{E}(\vec{x})$.  If all principal components are used (i.e. $d=p$) then it is easy to see that $\vec{g}\circ\vec{f}$ is the identity and the reconstruction error vanishes. 

A further difference between principal curves and NLPCA is that the projection $\vec{f}$ in NLPCA has to be continuous whereas the projection in principal curves can be discontinuous \cite{Malthouse:98}. An important concept for NLPCA as well as for principal curves is that of self-consistency, implying that the reconstructed curve is the mean over all data with equal projection, i.e $\vec{g}(\vec{\eta}) = \mathbb{E}(\vec{x}|\vec{f}(\vec{x})=\vec{\eta})$. This idea is the cornerstone of the original principal curve approach by  Hastie and Stuetzle \cite{HasStue:89} (hereafter: HS) as well as for the NLPCA by Bolton et al. \cite{Boletal:03},  who also consider all possible combinations of linear and nonlinear projection and  reconstruction. A comparison of NLPCA and HS principal curves can be found in \cite{Monaghan:00}.

A special type of principal curves,  a so-called {\it local principal curve} (\cite{lpc}, see Section \ref{sec:lpc}), is shown for the second of the two data sets in figure \ref{dez710}.

This article is concerned with extensions of PCA based on {\it localization}. 
 In Section 2 we review such methods and also present a new approach based on local partitioning. In Section 3 we discuss how principal component regression can be localized. Furthermore we discuss projection directions other than the principal components. Using the projection directions used in the partial least squares (PLS) algorithm allows for taking both the inherent structure of the covariates and the response variable into account. In Section 4 we apply these methods to simulated data produced in order 
to develop classification and parametrization methods for the upcoming 
Gaia astronomical survey mission. This survey will obtain spectroscopic 
data (i.e. high dimensional vectors) on over one billion ($10^9$) stars, 
from which we wish to estimate intrinsic astrophysical parameters 
(temperature, abundance, surface gravity).

%% file: sec2.tex
\section{Localized principal component analysis}

The term ``local(ized) PCA''  was coined decades ago in the literature on dimension reduction and feature extraction.  However, it has not been used in a unique way; there are several ways of interpreting  the term \textsl{local(ized)}. In this section, we will give a brief review of types of local PCA and illustrate some of them by applying them to the second speed-flow diagram presented in Section 1.  In addition, we consider another artificial data set (corresponding to figure 3d in \cite{lpc}), representing a spiral-like data cloud, with which principal curve / NLPCA algorithms generally tend to have problems.

\subsection{Cluster-wise PCA}

Historically the term \textsl{localized PCA} was used for cluster-wise PCA. Instead of fitting principal components through the whole data set, cluster-wise PCA partitions the data cloud into clusters, within which the principal component analysis is carried out ``locally''.   

Originally cluster-wise PCA was proposed as a tool for exploratory data analysis \cite{fugu, Braverman:70}. It was rediscovered by the neural network community in the 1990s in the context of non-linear signal extraction as an alternative to auto-associative neural networks \cite{KamLeen:97}. Five-layer auto-associative neural networks were successfully employed for (global) nonlinear PCA \cite{oja91}, using input and output layers with $p$ units, and a hidden layer with $d<p$ units. Auto-associative neural networks possess many desirable theoretical properties in terms of signal approximation \cite{horny}. However neural networks typically yield a non-convex optimization problem that is burdensome to solve, and the algorithm is --- without suitable regularization --- prone to getting trapped in poor local optima. Empirical evidence \cite{KamLeen:97} suggests that cluster-wise PCA is at least on a par with auto-associative neural networks.

The cluster-wise PCA algorithm can be seen as a generalization of the Generalized Lloyd algorithm (``k-means'') \cite{ggkm}. In the k-means algorithm the cluster centers are points, whereas in localized PCA, the cluster centers are hyperplane segments. The outline of the the cluster-wise PCA algorithm is as follows \cite{Diday:79}: 

\vspace{0.2cm}
\noindent {\bf Algorithm 1: Cluster-wise (``local'') PCA} \hrulefill
\begin{enumerate}[1.]
\item Choose a target dimension $d$ ($d=1$ e.g. yields a line), a number of clusters $Q$, and an initial partitioning of the input space into $Q$ disjoint regions $R^{(1)}\cup\ldots\cup R^{(Q)}=\mathbb{R}^p$.
\item Iterate \ldots
\begin{enumerate}[i.]
\item For each partition $R^{(q)}$ ($q=1,\ldots,Q$) compute the ``local'' covariance matrices
\[
 \vec{\Sigma}^{(q)} = \frac{1}{n^{(q)}}\sum_{i\in R^{(q)}}(\vec{x}_i-\bar{\vec{x}}^{(q)})(\vec{x}_i-\bar{\vec{x}}^{(q)})'\] with $ 
\bar{\vec{x}}^{(q)}=\frac{1}{n^{(q)}}\sum_{i\in R^{(q)}}\vec{x}_i
$ and $n^{(q)}$ being the number of observations in partition $R^{(q)}$.
Obtain the local principal components $\vec{\gamma}_1^{(q)},\ldots, \vec{\gamma}_d^{(q)}$ from the eigen decomposition of $\vec{\Sigma}^{(q)}$ (with corresponding eigenvalues $\lambda_1^{(q)} \geq \ldots \geq \lambda_d^{(q)} \geq \ldots$). 

\item Update the partitioning $R^{(1)},\ldots,R^{(Q)}$ by allocating each observation $\vec{x}_i$ to the ``nearest'' partition, i.e. the partition whose hyperplane segment is closest to $\vec{x}_i$.
\end{enumerate}
\end{enumerate}
\hrule
\vspace*{0.5cm}

Note that it is important in step 2.ii. to only consider the hyperplane segment  $\left \{\bar{\vec{x}}^{(q)}+\sum_{j=1}^d \xi_j \vec{\gamma}_j^{(q)}:\; \xi_1,\ldots\xi_d\in \mathbb{R}\right\} \cap R^{(q)}$ instead of the whole hyperplane.  
Originally the partitioning was fixed a priori and thus there was no need to iterate steps i. and ii. This much simpler setup however is typically detrimental to the performance of the algorithm. A variant of algorithm 1 featuring automatic selection of clusters, and of the number of principal components within clusters, was proposed by Liu et al. \cite{Liuetal:03}  and used for star/galaxy classification. 

It is easy to see that in complete analogy to the k-means algorithm the sum of squared distances (\ref{XminEmp}) is never increased by any step of the above algorithm. However as with k-means or auto-associative neural networks, there is no guarantee that the global optimum is found. The most important drawback of this method is that its results are highly dependent on the initial choice of the partitioning.

This gives rise to the idea of slowly ``building up'' the partitions recursively akin to classification and regression trees (CARTs) \cite{Breiman.84}. Starting with a single partition, each partition is recursively ``split up'' into two partitions if the data can locally be better be approximated by two hyperplane segments instead of a single one. 

In order to evaluate whether such a split is necessary, every partition $R^{(q)}$ is split at the mean of the partition  $\bar{\vec{x}}^{(q)}$ orthogonally to the first principal component $\vec{\gamma}^{(q)}$ of the data in partition $q$, yielding two partitions $R^{(l)}$ and $R^{(r)}$.  Next a small number of ``k-segments'' steps (steps 2.i. and 2.ii. from algorithm 1) are carried out for the partitions $R^{(l)}$ and $R^{(r)}$, however only using the data initially belonging to the partition $R^{(q)}$. The split is retained if   
\begin{equation}\label{ludcrit}
\frac{\lambda^{(q)}_1+\ldots+\lambda^{(q)}_d}{\lambda^{(q)}_1+\ldots+\lambda^{(q)}_p} < C \cdot \left( \frac{n^{(l)}}{n^{(q)}} \cdot \frac{\lambda^{(l)}_1+\ldots+\lambda^{(l)}_d}{\lambda^{(l)}_1+\ldots+\lambda^{(l)}_p} + \frac{n^{(r)}}{n^{(q)}} \cdot \frac{\lambda^{(r)}_1+\ldots+\lambda^{(r)}_d}{\lambda^{(r)}_1+\ldots+\lambda^{(r)}_p}\right),
\end{equation}
where $n^{(q)}$ is the number of observations in partition $R^{(q)}$, $\lambda_k^{(q)}$ the $k$-th largest eigenvalue of the covariance of the data in partition $R^{(q)}$, and $C$ is a constant which is typically chosen to be $1$. Note that, contrary to CARTs, only a single split is considered for each partition. Considering every possible split of the partition would be computationally wasteful and in most cases not necessary as the split is subsequently optimized using a few ``k-segments'' steps.

In addition to testing whether partitions should be split, one can test whether neighboring partitions should be joined. This allows the algorithm to get around some of the suboptimal local extrema. In our experience it is beneficial to use a rather loose definition of neighborhood. Two partitions $R^{(l)}$ and $R^{(r)}$ are hereby considered to be neighbors if for at least for one observation $\vec{x}_i$, both $R^{(l)}$ and $R^{(r)}$ are amongst the $s>2$ ``closest'' partitions. If the inequality (\ref{ludcrit}) does not hold for two neighboring partitions $R^{(l)}$ and $R^{(r)}$, these are joined forming a new partition $R^{(q)}=R^{(l)}\cup R^{(r)}$.

This yields the following new algorithm:

\vspace{0.2cm}
\noindent {\bf Algorithm 2: Recursive local PCA} \hrulefill
\begin{enumerate}[1.]
\item Start with a single partition $R^{(1)}$ containing all the data.
\item Iterate \ldots
\begin{enumerate}[i.]
\item Test for each partition $R^{(q)}$ whether it should be split (using the criterion (\ref{ludcrit})).
\item Carry out a fixed number of ``k segments'' steps (steps 2.i. and 2.ii. of algorithm 1) updating all partitions.
\item Test for each neighboring pair of partitions whether the pair should be joined (using the criterion (\ref{ludcrit})).
\end{enumerate}
\dots until there is no change in the allocation of observations to partitions.
\end{enumerate}
\hrule
\vspace*{0.5cm}

Note that this algorithm does not require the choice of an initial partitioning. It can be beneficial to ``enforce'' a certain number of splits during the first few iterations, i.e. initially carry out every possible split irrespective of whether it meets criterion (\ref{ludcrit}) and initially skip step 2.ii.  

A similar idea has been proposed by Verbeek et al. \cite{verbeek00ias}. Their method is however based on splitting off ``zero length'' segments corresponding to each observation.

\begin{figure}
 \hspace*{-0.1cm}\includegraphics[scale=0.6, angle=-90]{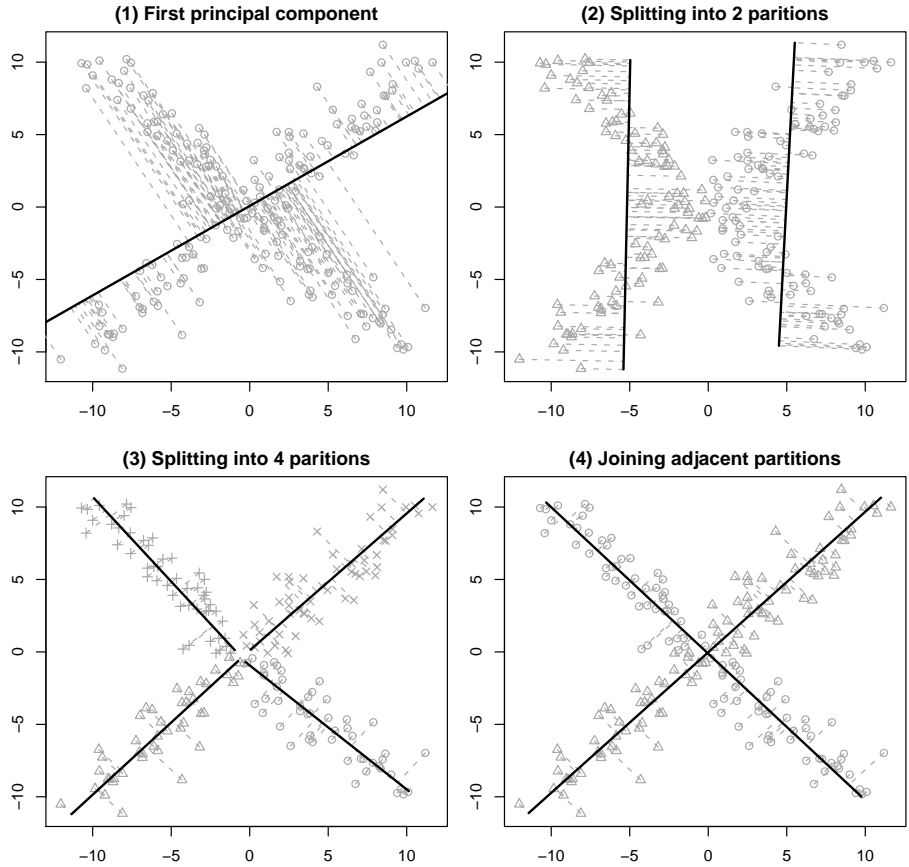}
         \caption{\label{ludex} Example illustrating the recursive local PCA algorithm.}
\end{figure}

Figure \ref{ludex} illustrates the algorithm using a simple example. The data cannot be summarized by its first principal component (top left panel). Whilst it can be summarized by two line segments, the two line segments found in the second step are no truthful representation of the data (top right panel). The four line segments found in the third step correspond to the structure behind the data (bottom right panel). Finally the adjacent segments are joined leading to two partitions only (bottom right panel).

Note that the algorithm does not yield a principal curve or manifold, but merely disconnected line or hyperplane segments. It can be seen as finding tangent approximations to the principal curve or manifold. Thus the algorithm can by design cope with disconnected or branching principal curves or manifolds. Note that in the case of principal curves ($d=1$) one can join the line segments to build a polygonal line \cite{verbeek00ias}.

\subsection{Principal curves}\label{sec:lpc}

Principal curves were firstly introduced by Hastie and Stuetzle \cite{HasStue:89}.
 Their definition is based on the concept of self-consistency. A curve $\vec{g}$ is called a Hastie-Stuetzle (HS) principal curve if  $\vec{g}(\eta)= \mathbb{E}(\vec{x}|\eta_{\vec{g}}(\vec{x})=\eta)$, with projection index  $\eta_{\vec{g}}(\vec{x}):=\arg\min_{\eta}\|\vec{x}-\vec{g}(\eta)\|$. Hastie and Stuetzle show that a HS principal curve is a critical point of the distance function (\ref{Xmin}). Hastie and Stuetzle's definition however has a number of shortcomings. Whilst the principal curve can be shown to be a critical point of the distance function, one can show under fairly general conditions that it is just a saddle point of the distance function \cite{dustu1}. Further, when the data $\vec{x}$ are generated by adding noise $\vec{\epsilon}$ with $\mathbb{E}\vec{\epsilon}=\vec{0}$ to a curve $\vec{g}$, i.e. $\vec{x}=\vec{g}(\eta)+\vec{\epsilon}$, $\vec{g}$ is typically not the principal curve.  
K\'egl et al. \cite{Kegl-etal:00} overcome the former problem by considering principal curves of fixed length. Tibshirani's definition of principal curves \cite{Tibshirani:92} overcomes the latter shortcoming by defining principal curves using a generative model.

The algorithms proposed by Hastie and Stuetzle, Tibshirani, and K\'egl can all be described as ``top-down'' (or ``global'') algorithms. All start with the first principal component and then iteratively ``bend'' it. The algorithm by Hastie and Stuetzle  can be summarized as follows.

\vspace{0.2cm}
\noindent {\bf Algorithm 3: ``Top-down'' principal curves (HS)} \hrulefill
\begin{enumerate}[1.]
\item Initialize the principal curve as the first principal component line $\vec{g}(\eta)=\bar{\vec{x}} + \eta\vec{\gamma}_1$.
\item Iterate between \dots
\begin{enumerate}[i.]
\item {\bf Projection} Project the data points $\vec{x}_i$ onto the principal curve $\vec{g}$ yielding projection indexes $\eta_{\vec{g}}(\vec{x}_i)= \eta_i$.
\item {\bf Reconstruction} Fit the data points  $\vec{x}_i$  component-wise against the projection indices $\eta_i$ using a scatterplot smoother. 
\end{enumerate}
\dots  until the change of some measure of goodness-of-fit falls below a certain threshold. The curve $\vec{g}:\RR \longrightarrow \RR^p$ reconstructed in the last iteration is the estimated principal curve.
\end{enumerate}
\hrule
\vspace*{0.5cm}

The reconstruction step requires some form of smoothing, otherwise all data points would just be interpolated. It is implemented by using splines or local smoothers such as LOESS \cite{Cleveland:79}. Note that this algorithm entails that the order of the projections $\eta_i$ is maintained in each iteration.   Tibshirani's approach yields a similar algorithm. As Tibshirani's definition does not assume errors that are orthogonal to the principal curve the observations cannot be directly associated to a single point on the curve. Rather the algorithm is based on weights (``responsibilities'') that correspond to the likelihood that a certain observation was generated by a certain point on the curve. From this point of view Tibshirani's algorithm is very similar to the EM algorithm for Gaussian mixture models. K\'egl et al.'s algorithm uses polygonal lines with an increasing number of nodes. 

\begin{figure}
\centerline{\hspace*{-0.5cm}\includegraphics[angle=270,width=12.2cm]{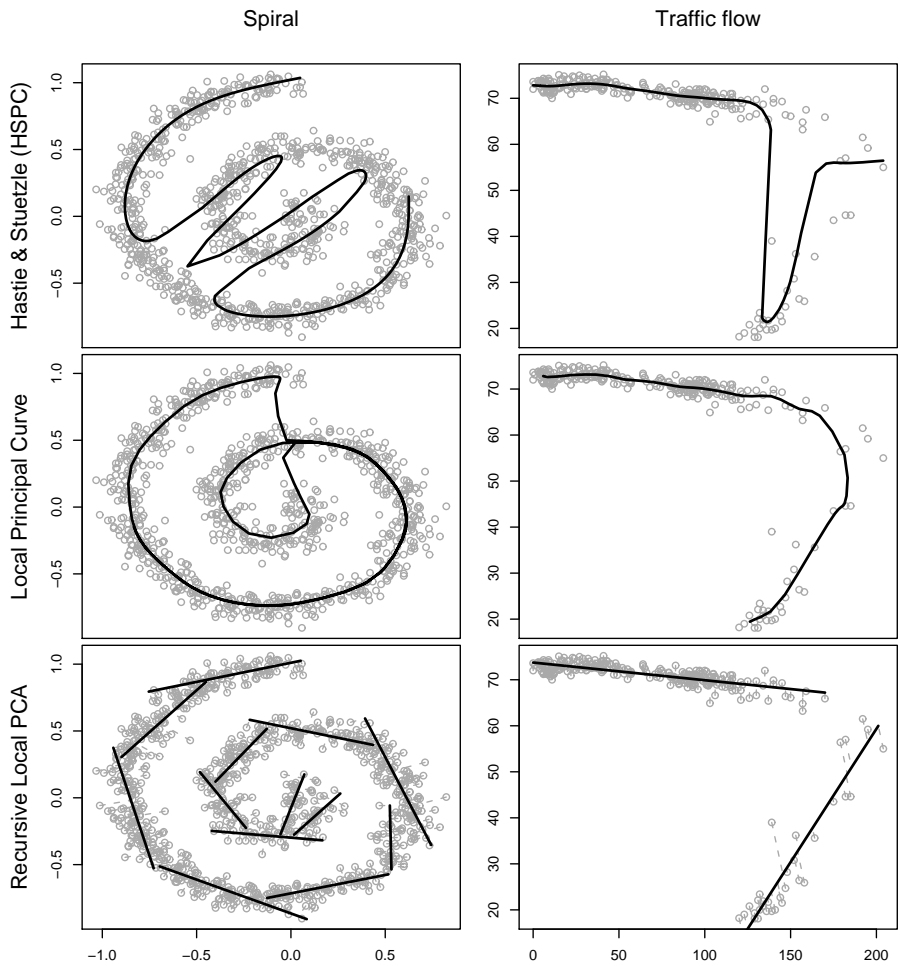}}
\caption{\label{beispiele} Principal curves obtained for the spiral data and the non-linear traffic flow data using algorithm 3 (top row), algorithm 4 (middle row), and algorithm 2 (bottom row)}
\end{figure}

Though all the three algorithms were shown to give satisfying results in a variety of circumstances, they all suffer from the problem of being highly dependent on the initialization. An unsuitable initialization of the projection indices $\eta_i$ cannot be corrected later on. This is particularly obvious for spiral-like data, see figure \ref{beispiele} (top).

Hence, instead of starting with a {\it global} initial
line, it often seems more appropriate to construct the principal
curve in a ``bottom-up'' manner by considering in every step only
data in a {\it local} neighborhood  of the currently  considered
point.  Delicado \cite{Delicado:01}  proposed  the  first
principal curve approach which can be assigned to this family, the {\it principal curves of oriented points (PCOP)}.  A second approach in this direction
was recently made by Einbeck et al. \cite{lpc} and is known as {\it local principal curves (LPC)}. These two algorithms differ from all other existing NLPCA / principal curve algorithms in several crucial points:
\begin{itemize}
 \item They do not start with an initial globally constructed line like the first principal component.
 \item They do not dissect into the stages of projection and reconstruction. The entire fitting process is carried out in the data space only.
 \item They do not maximize/minimize a global fitting criterion.
\end{itemize}

We illustrate this family of methods by providing the LPC algorithm explictly. 
 To fix terms, let $K_{\vec{H}}(\cdot)$ be a $p-$dimensional kernel function, $\vec{H}$ a multivariate $(p \times p)$ bandwidth matrix,  $w_i^{\vec{x}}=K_{\vec{H}}(\vec{x}_i-\vec{x})/\sum_{i=1}^nK_{\vec{H}}(\vec{x}_i-\vec{x})$, and  $\vec{x}_i' =(x_{i1}, \ldots, x_{ip}),\, i=1, \ldots, n$.

\vspace{0.2cm}
\noindent {\bf Algorithm 4: Local principal curves (LPC)} \hrulefill
\begin{enumerate}[1.]
\item Select a starting point  $\vec{x}_0$ and a step size $t_0$. Set $\vec{x} = \vec{x}_0$. 
\item Calculate the local center of mass 
$\vec{\mu}^{\vec x}= {\sum_{i=1}^nw_i^{\vec{x}}\,\vec{x}_i}$
at $\vec{x}$. Denote by $\mu_j^{\vec x}$ the $j$-th component of $\vec{\mu}^{\vec x}$. 
\item Estimate the local covariance matrix $\vec{\Sigma}^{\vec x}=(\sigma_{jk}^{\vec x}) $ at $\vec{x}$ via
\begin{center}
$\sigma_{jk}^{\vec x}=\sum_{i=1}^nw_i^{\vec{x}}(x_{ij}-\mu_j^{\vec x})(x_{ik}-\mu_k^{\vec x})$.
\end{center}
Let $\vec{\gamma}^{\vec x}$ be the first column of the loadings matrix $\vec{\Gamma}^{\vec x}$ computed locally at $\vec{x}$ in analogy to (\ref{pcdecom}).
\item
Setting 
$\vec{x} := \vec{\mu}^{\vec x} + t_0\vec{\gamma}^{\vec x}$,
one finds the updated value of $\vec{x}$. 
\item
Repeat steps 2 to 4 until the sequence of $\vec{\mu}^{\vec x}$ remains approximately constant (implying that the end of the data cloud is reached).  Then set again  $\vec{x}=\vec{x}_0$, set $\vec{\gamma}^{\vec x} :=-\vec{\gamma}^{\vec x}$ and continue with step 4.
\vspace{0.2cm}
\hrule
\end{enumerate}

The step size $t_0$ is  recommended to be set equal to $h$ if $\vec{H}=\mbox{diag}(h, \ldots, h)$.
The starting point $\vec{x_0}$ may or may not be a member of the data cloud, and can be selected by hand or at random. Depending on the particular data set, it can have a rather crucial influence on the fitted curve. (In the examples provided in figure \ref{beispiele} (middle),  the spiral fit is quite independent of $\vec{x_0}$, but it may require some attempts to get the fit through the traffic data right. A successful (randomly selected) choice was here e.g. $\vec{x_0}= (93,71)$,  using $h=t_0=7$.) For branched or disconnected data clouds it is often useful to work with multiple initializations \cite{ETEgfkl:05,lpc} and to compose the principal curve from the individual  parts resulting from each starting point.
In order to improve stability it is beneficial to replace the first local principal component $\vec{\gamma}^{\vec{x}}$ by the weighted average between the current local principal component $\vec{\gamma}^{\vec{x}}$ and the previous local principal component $\vec{\gamma}^{\vec{x}_{\textrm{old}}}$, i.e. use $\alpha \vec{\gamma}^{\vec{x}} + (1-\alpha)  \vec{\gamma}^{\vec{x}_{\textrm{old}}}$ for a suitable $\alpha \in (0,1]$ (``angle penalization'', see \cite{lpc}).

To make the difference to Delicado's algorithm clear, let $\mathcal{H}(\vec{x},\vec{b})$ be the hyperplane which contains $\vec{x}$ and is orthogonal to a vector $\vec{b}$. Then, for given $\vec{x}$, Delicado defines the {\it principal direction} $\vec{b}^*(\vec{x})$ as the vector minimizing the total variance of all observations lying on this hyperplane, and $\vec{\mu}^*(\vec{x})$ as the expectation of all points lying on $\mathcal{H}(\vec{x},\vec{b}^*(\vec{x}))$. The set of fixed points of $\vec{\mu}^*$ defines the set $\mathcal{P}(\vec{X})$ of {\it principal oriented points (POPs)}, and any curve $\alpha: \mathbb{R} \supset I \longrightarrow \mathbb{R}^p$ with $\{\alpha(s): s \in I\} \in \mathcal{P}(\vec{X})$ is a PCOP. 
This is a probabilistic definition, hence estimates are required in practice, and this is where localization enters into Delicado's algorithm in two different ways. Firstly, %in order to estimate $\vec{\mu}^*(\vec{x})$ 
one considers a certain neighborhood of the current hyperplane $\mathcal{H}$, and data nearer to $\mathcal{H}$ are associated with higher weights. Secondly, a cluster analysis is performed on $\mathcal{H}$, and only data in the local cluster are used for averaging.  In contrast to the LPCs, where the local center of mass is found in a one-step approximation, Delicado's analogue to step 2 requires iteration until convergence. Summarizing the essential difference slightly simplified, whereas LPC is based on calculating alternately a local first principal component $\vec{\gamma}_1^x$ and a local center of mass $\vec{\mu}^x$, in  Delicado's algorithm these two components are replaced by the estimated principal direction and the estimated fixed points (POPs), respectively. While Delicado's algorithm is based on an elaborated and sound probabilistic theory, it is quite burdensome from a computational point of view, as his principal directions are not obtainable through an eigen decomposition \cite{lpc}, and a large number of cluster analyses has to be run. On the other hand, the computationally faster  LPCs are a purely empirical concept, but they can be shown to be an approximation of Delicado's algorithm for small cluster sizes \cite{lpc}.  For either algorithm, extensions to branched principal curves using the notion of {\it local second principal curves} have been suggested \cite{Delicado:01,ETEgfkl:05} and automatic bandwidth selection routines have been proposed \cite{DelHue:03,lpc}.

\subsection{Further approaches}

The term local PCA has also been used for other techniques and in other contexts,  which we briefly summarize here.  
Firstly, one can extend the algorithms presented in Section 2.1 by allowing for smoothness over clusters  using mixture models \cite{Marchette:98}. This means, rather than allocating an observation  $\vec{x}_i$ to the nearest partition $q$, one considers the posterior probabilities $\vec{\pi}_{i,q}$ that  $\vec{x}_i$ is generated by partition $q$, and defines {\it for each observation} the weighted (first) local principal component $\vec{\gamma}_{1,i}= \sum_q \vec{\pi}_{i,q}\vec{\gamma}_1^{(q)}$. This method requires knowledge of the mixture density itself and produces eigenvectors which are biased towards adjacent components, which motivated \cite{Marchette:98} to provide a modified algorithm using Fisher's linear discriminant (FLD) instead of PCA.

Secondly, there are approaches from the NLPCA family using local methods. While most NLPCA algorithms work by applying some nonlinear, but parametric, transformation in the projection and/or the reconstruction step (\cite{Webb:96}, \cite{ChaGi:99}, among others),   Bolten et al. \cite{Boletal:03} allowed for a nonparametric reconstruction $\vec{g}$. Using in the projection step  a nonlinear transformation followed by a linear mapping, i.e. $\vec{f}(\vec{x})= {\vec W}\vec{\phi}(\vec{x})$, $\vec{\phi}:\mathbb{R}^p \rightarrow \mathbb{R}^l, \vec{W} \in \mathbb{R}^{d \times l}$, the reconstructed curve takes the form $\vec{g}({\vec W}\vec{\phi}(\vec{x}))$, which is estimated  using projection pursuit regression \cite{friedppr}.  The actual smoothing method employed is the so-called supersmoother \cite{Friedmann:84}, which is a running line smoother using a local neighborhood of the target point with variable span. 
We do not consider NLPCA approaches further, as the contribution by U. Kruger in this volume is explicitly dedicated to them. 

Thirdly, the term ``local principal component analysis'' was used by Aluja-Banet and Nonell-Torrent \cite{AjulaNonell:91} for PCA using contiguity relations based on a non-directed graph expressing binary relations between the individuals. Their motivation is to control for the effect of a latent or third variable (e.g. geographical position)  by virtually eliminating its influence on the principal component analysis. This falls somewhat out of the context of the present paper and we do not consider it further.

%% file: sec4.tex
\section{Combining principal curves and regression}

\subsection{Principal component regression and its shortcomings}
Using principal components for dimension reduction prior to carrying out further analysis is an old paradigm in statistics. The combination of principal component analysis and linear regression, known as principal component regression (PCR), was proposed already in the 1950s by Kendall \cite{kendall} and Hotelling \cite{hotelling}. PCR consists of two steps. As a first step the first $d$ principal components are extracted from the covariates $\vec{X}$. In the second step, the first $d$ principal component scores $\vec{T}_d:=\left(\vec{t}_1\cdots \vec{t}_d\right)=\vec{X}\cdot\left(\vec{\gamma}^{(1)}\cdots\vec{\gamma}^{(d)}\right)$ are used as covariates in a linear regression model. Geometrically, this corresponds to projecting the covariates onto the affine hyperplane spanned by the first $d$ principal components and using the projection indices as covariates in the regression.

The purpose of the projection step is to eliminate directions in which the covariates have little variance. The rationale behind this is that directions with little variance correspond to directions in which there is little information in the covariates and thus are prone to leading to an overfit.

Being built upon principal component analysis, PCR suffers from the same shortcomings as PCA. If the structure of the covariates cannot be suitably approximated by an affine hyperplane of low dimension, PCA and thus PCR are likely to fail.
In the following we will propose a new regression-tree like algorithm, based on algorithm 2, which  can be used for local dimension reduction of the covariate space of a regression model.

\subsection{The generalization to principal curves}
Instead of projecting the covariates onto the principal component surface we can project them onto a principal curve or manifold. In complete analogy to PCR we can then use the projection indices in the regression model. 

This requires a unique parameterization of the principal manifold (e.g. a suitable unit speed parameterization). Whilst some of the algorithms presented in Section 2 compute the projection indices, other principal curve or manifold algorithms like algorithm 4 only yield a set of points on the curve or manifold. 

The problem of having to compute the projection indices onto a non-linear curve or manifold can be circumvented by using tangent approximations to the principal curve or manifold as provided by algorithm 2, because the projections are then onto hyperplane segments and not onto non-linear manifolds. The projections onto the hyperplane segment corresponding to partition $R^{(q)}$ are
$$(t^{(q)}_{i1},\ldots,t^{(q)}_{id})=\mathop{\arg\min}_{\stackrel{t^{(q)}_{i1},\ldots,t^{(q)}_{id}\in\mathbb{R}}{\bar{\vec{x}}^{(q)}+\sum_{j=1}^d t^{(q)}_{ij}\vec{\gamma}^{(q)}_j\in R^{(q)}}}
\left\|\vec{x}_i-\bar{\vec{x}}^{(q)}-\sum_{j=1}^d t^{(q)}_{ij}\vec{\gamma}^{(q)}_j
\right\|^2.
$$
The $(t^{(q)}_{i1},\ldots,t^{(q)}_{id})$ are easy to compute as they are either the orthogonal projection of the covariates onto the hyperplane or a point on the boundary of the hyperplane segment.

In analogy to algorithm 2 one can fit a regression model separately in each partition. This, however, leads to a couple of disadvantages. First of all, the prediction would be discontinuous at the boundaries between the partitions. Furthermore, keeping the ``hard partitioning'' can lead to a large variance of the predictions, especially if the number of partitions is large. In order to avoid these detrimental effects the partitioning is ``softened'' by the introduction of weights that reflect how far an observation $\vec{x}_i$ is from a certain hyperplane segment. In order to achieve this all observations are projected onto all of the hyperplane segments. For each partition $R^{(q)}$ weights $\omega_i^{(q)}$ are associated with each observation $\vec{x}_i$. The weight $\omega_i^{(q)}$ decreases with increasing distance between the observation and the hyperplane, i.e. $
\omega^{(q)}_i:=w({\delta^{(q)}_i}) 
$,
where $\delta_i^{(q)}=\left\|\vec{x}_i-\bar{\vec{x}}^{(q)}-\sum_{j=1}^d t^{(q)}_{ij}\vec{\gamma}^{(q)}_j
\right\|^2$ is the squared distance between the $i$-th observation and its projection onto the hyperplane segment belonging to the partition $R^{(q)}$, and $w:\mathbb{R}^+_0 \rightarrow \mathbb{R}^+_0$ is a decreasing function, such as $w(\delta)=e^{-C\cdot \delta}$. Figure \ref{ludger_illus} illustrates this idea. The steps of the proposed algorithm are as follows:

\begin{figure}
\centerline{\includegraphics[height=4.5cm]{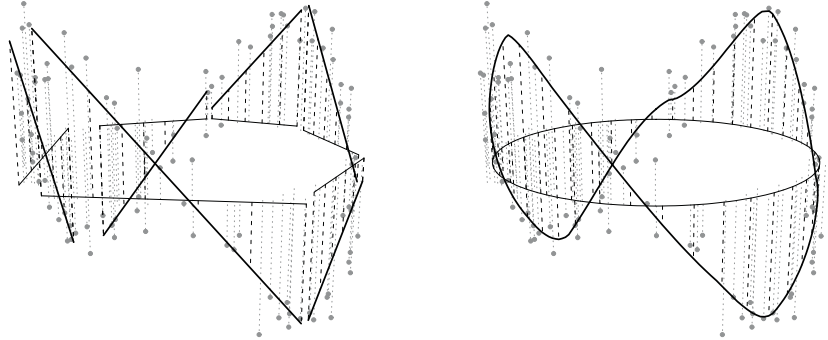}}
\caption{Local regression lines in the partitions and resulting regression function\label{ludger_illus} for covariates lying on a circle in the two-dimensional plain. The vertical dimension corresponds to the response. Left: ``hard partitioning''; right: ``softened''.}

\end{figure}

\vspace{0.2cm}
\noindent {\bf Algorithm 5: Projection-based regression trees} \hrulefill
\begin{enumerate}[1.]
\item Carry out algorithm 2 yielding a set of partitions $R^{(1)},\ldots,R^{(Q)}$ with the corresponding hyperplane segments.
\item Compute the squared distances $\delta_i^{(q)}=\left\|\vec{x}_i-\bar{\vec{x}}^{(q)}-\sum_{j=1}^d t^{(q)}_{ij}\vec{\gamma}^{(q)}_j
\right\|^2$ and weights $\omega^{(q)}_i:=w({\delta^{(q)}_i})$.
\item For each partition $R^{(1)},\ldots,R^{(Q)}$:
\begin{itemize}
\item[] Fit a regression model using the projections of the data $\vec{T}^{(q)}:=\left(\vec{t}^{(q)}_1\cdots \vec{t}^{(q)}_d\right)$ as covariates, $\vec{y}$ as response, and $\omega^{(q)}_i$ as observation weights.
\end{itemize}
\item Compute the predictions as weighted means 
$$
\hat y_i=\frac{\sum_{q=1}^Q \omega^{(q)}_i \hat y_i^{(q)}}{\sum_{q=1}^Q \omega^{(q)}_i},
$$
where $\hat y_i^{(q)}$ is the prediction obtained from the regression model in partition $R^{(q)}$.
\end{enumerate}
\hrule
%\vspace*{0.2cm}

Note that a wide variety of regression methods can be used in step 3. If the regression method does not support weights one can alternatively sample from all observations using the $\omega^{(q)}_i$ as sampling weights.

\begin{figure}\begin{center}
\subfigure[``Classical'' CART]{\includegraphics[width=5.5cm,angle=270]{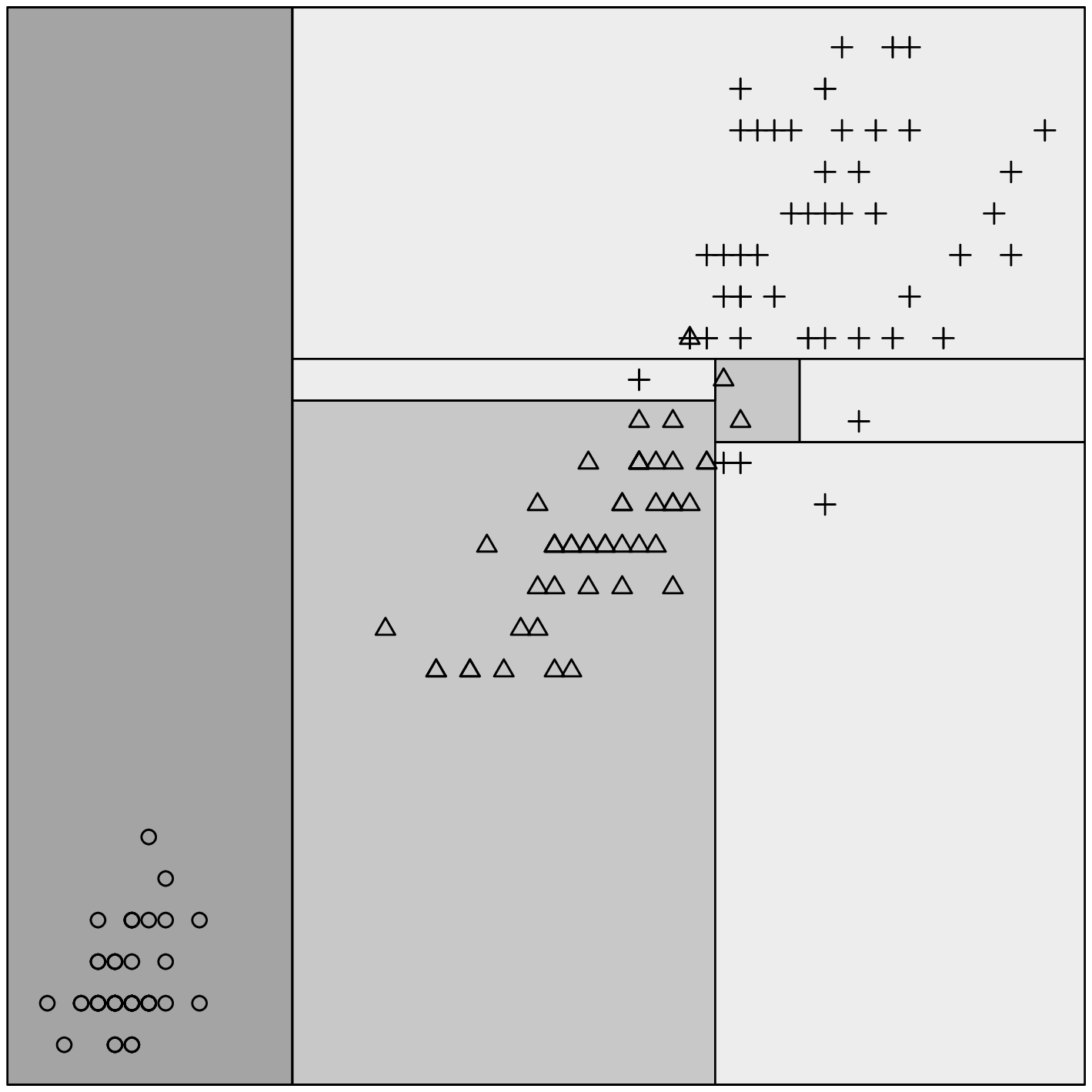}}
\hspace{0.2cm}
\subfigure[Projection-based classification tree]{\includegraphics[width=5.5cm,angle=270]{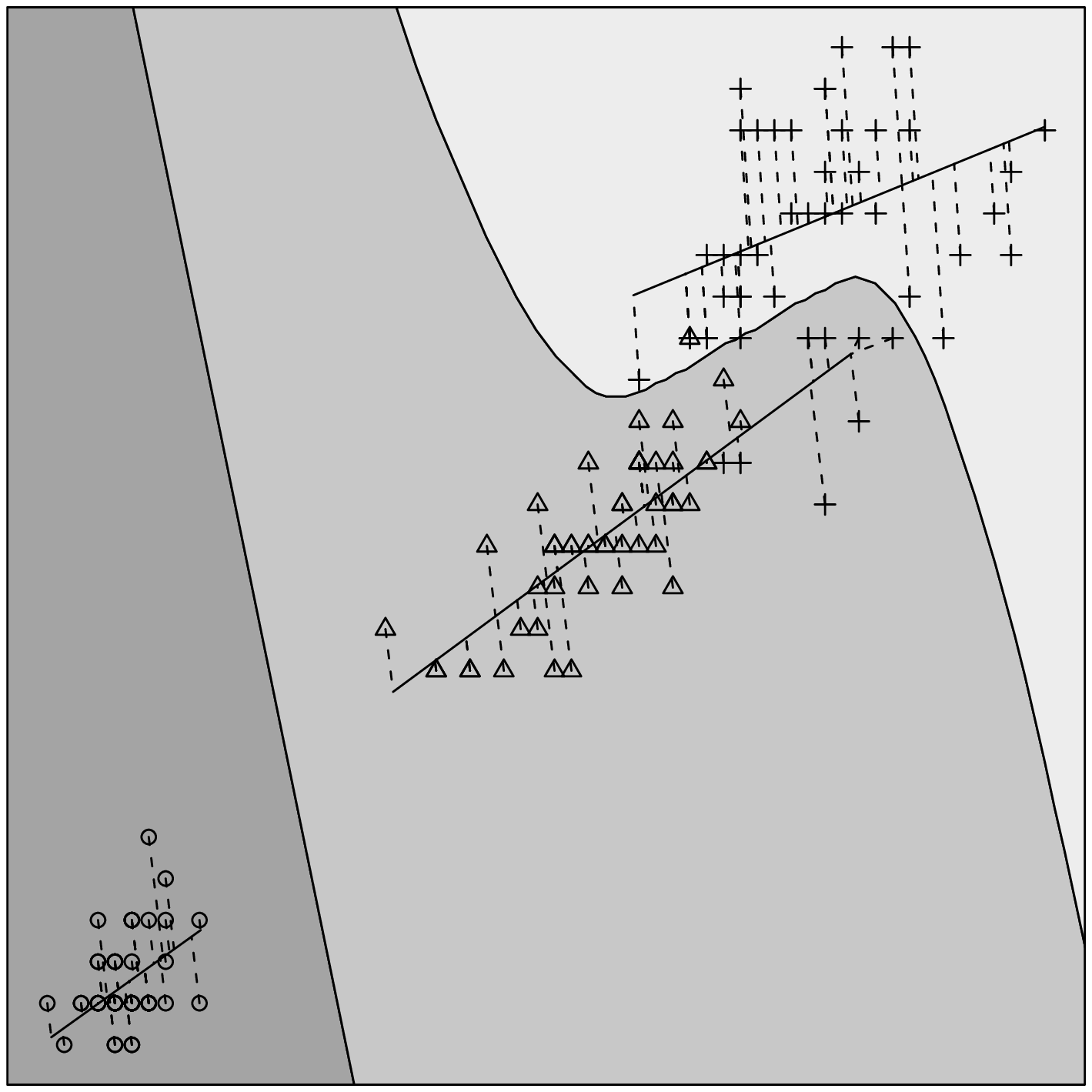}\vspace*{0.1cm}}
\caption{Comparison between a classification tree (CART) and a projection-based classification tree\label{ludger_illus2}}
\end{center}\end{figure}

Algorithm 5 is a prediction algorithm that is like CARTs based on recursive partitioning. In contrast to CARTs, which split up the covariate space by a series of (orthogonal) ``cuts'', the method proposed here partitions the covariate space according to a tessellation defined by the hyperplane segments. Figure \ref{ludger_illus2} illustrates this comparison. As algorithm 5 takes the structure of the covariates into account, it provides some sort of regularization. Thus it tends to have a smaller variance than CARTs, however at the price of a potentially increased bias. In a boosting \cite{freund96experiments} context, this makes the method proposed here an interesting candidate for a weak learner. 

The algorithm presented here is similar to cluster-wise linear regression \cite{spaeth}, however there a few major differences. Most importantly, a dimension reduction step is carried out in each cluster and the clusters are not constructed around a single point, the cluster center, but around hyperplane segments. Furthermore the boundaries between the partitions are softened by the exponential weights. Note that the algorithm considered so far does not take the response variable $\vec{y}$ into account when constructing the partitions.

\subsection{Using directions other than the local principal components}
The rationale behind projecting the data onto the principal component plane or a principal manifold was based on equating the variance of the covariates to the information the covariates provide. But this does not necessarily imply that the directions with the largest variance necessarily provide the \emph{relevant} information necessary to predict the response $\vec{y}$.\footnote{See \cite{jolliffe82} for a more detailed discussion.} In other words, it might be beneficial to consider both the structure of the covariates and the response variable.

We propose to modify algorithm 2 to take both these aspects into account. The core step of algorithm 2 were the ``k-segments steps'' used already in algorithm 1. In step 2.i. the principal components are computed in each partition. The core property of the principal components is that they maximize the variance of the projections $\mbox{Var}(\vec{X}^{(q)}\vec{\gamma})$. Clearly, this does not take the response $\vec{y}^{(q)}$ into account at all.\footnote{$\vec{X}^{(q)}$ contains hereby all covariates  $\vec{x}^{(q)}$ from the partition $R^{(q)}$, i.e. the $\vec{x}_i\in R^{(q)}$. $\vec{y}^{(q)}$ is defined analogously.}

If we were to choose the direction that predicts the response $\vec{y}^{(q)}$ \linebreak[4] best, we would choose the least squares regression estimate \linebreak[4]  $\hat{\vec{\beta}}^{(q)}={({\vec{X}^{(q)}}'\vec{X}^{(q)})}^{-1}{\vec{X}^{(q)}}'\vec{y}^{(q)}$, which maximizes the squared correlation coefficient $\rho^2(\vec{y}^{(q)},{\vec{X}^{(q)}}'\vec{\beta})$. Note that the latter criterion does not consider the variance of the covariates $\vec{X}^{(q)}$ at all.

It seems to be a suitable compromise to choose a projection direction $\vec{\gamma}$ that maximizes the product of the two aforementioned criteria, which is equivalent to maximizing the squared covariance between the $\vec{X}^{(q)}\vec{\gamma}$ and $\vec{y}^{(q)}$, i.e. 
$$
\mbox{Cov}^2(\vec{y}^{(q)},\vec{X}^{(q)}\vec{\gamma})=\rho^2(\vec{y}^{(q)},\vec{X}^{(q)}\vec{\gamma})\cdot \mbox{Var}(\vec{X}^{(q)}\vec{\gamma}) \cdot \mbox{Var}(\vec{y}^{(q)}).
$$ This covariance is maximised by the projection direction of the PLS algorithm \cite{wold1,simpls}. This alternative projection direction can easily be incorporated into algorithms 1 and 2. Step 2.i. of algorithm 1 simply has to be replaced by the computation of the PLS projection direction obtained by carrying out a PLS regression with the data belonging to partition $R^{(q)}$. Using this modification typically improves the predictive performance. The following paragraph gives an example for this.

\subsection{A simple example}
\begin{figure}\begin{center}
\hspace*{-1cm}
\includegraphics[width=6.5cm]{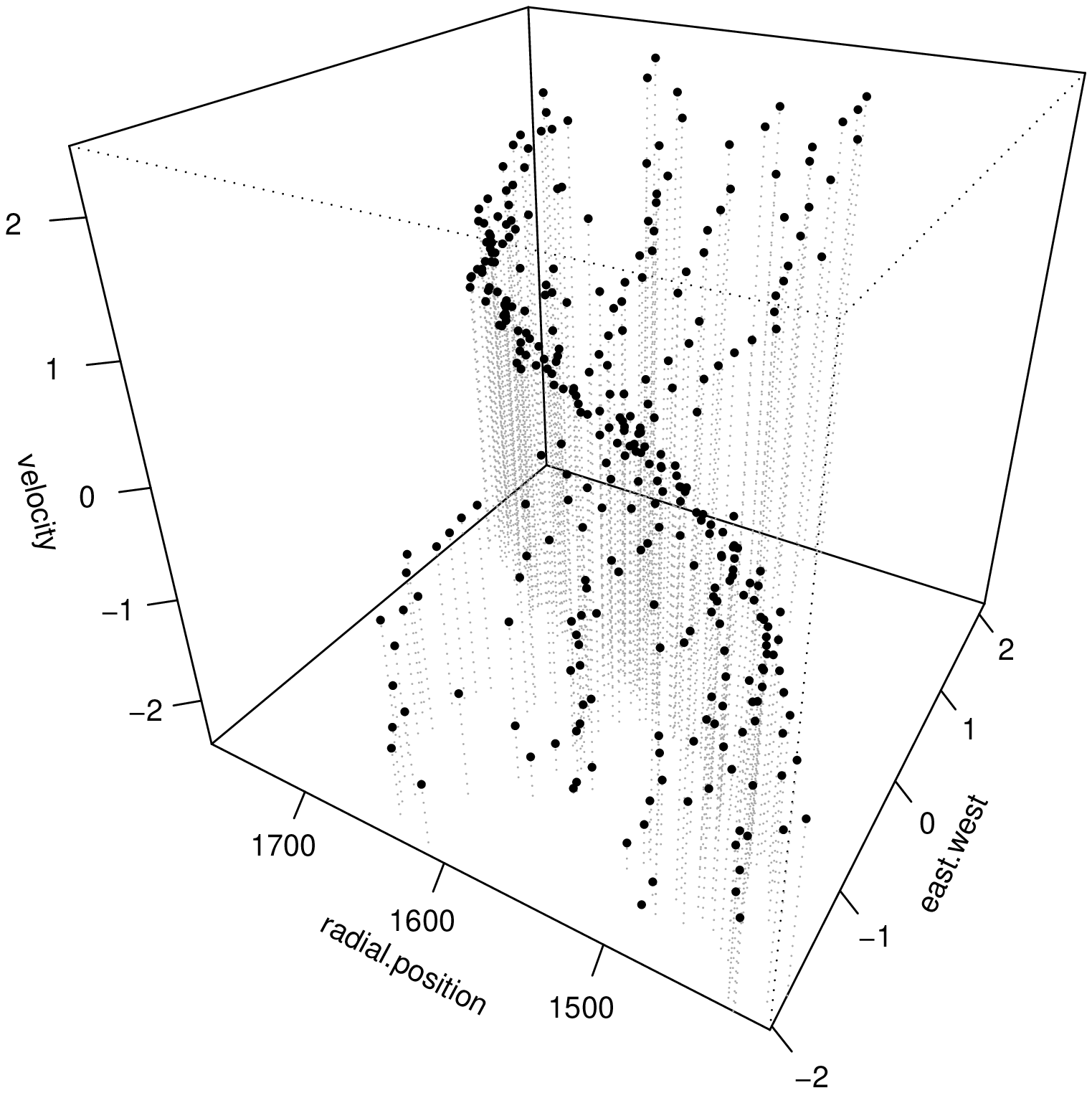}
\hspace*{-0.5cm}
\includegraphics[width=6cm]{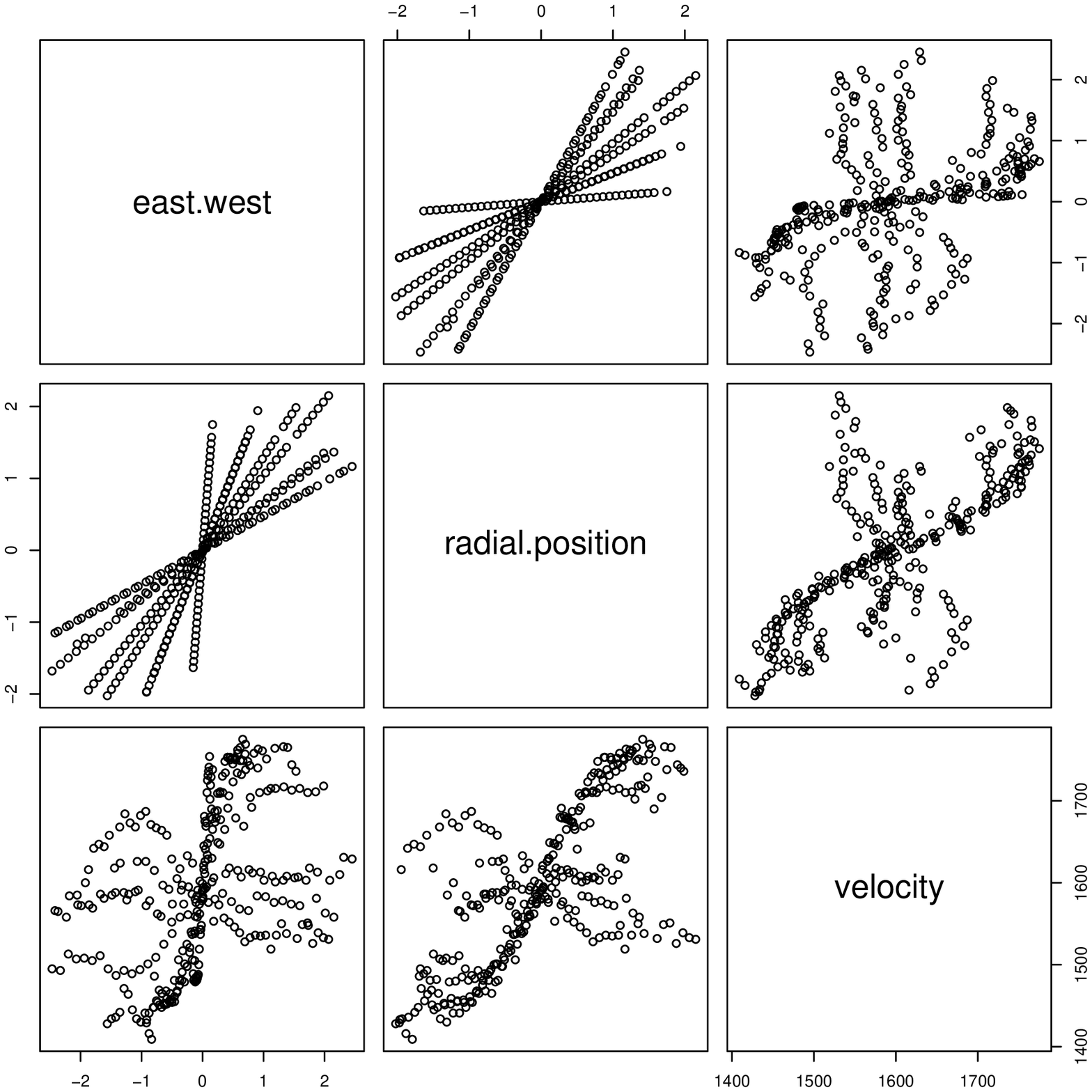}
\caption{\label{galaxy_data} Visualization of the galaxy data.}
\end{center}\end{figure}

In the following we will consider a data set taken from  \cite{sms}
to illustrate the algorithm proposed in this section.  The data consists of measurements of the radial velocity of  a spiral  galaxy  taken  at 323 points in the area of sky which it covers. The different points in the area of the sky are determined by their north-south and east-west coordinates. The data is visualized in figure \ref{galaxy_data}. It is easy to see that the measurements lie within a small number of slots crossing the origin. 

The data set was randomly split into a training set of 162, and a test set of 161 observations. Table \ref{galaxy_table} gives the $L_2$ error (and its standard deviation) obtained when predicting the radial velocity using different methods based on 1000 replications. The methods considered were algorithm 5 once with the principal components and once with the PLS projection direction, a linear model, generalized additive models (GAM) \cite{hastietibs}, multivariate adaptive regression splines (MARS) \cite{friedmars}, and projection pursuit regression (PPR) \cite{friedppr}. The results show that the algorithm proposed here clearly outperforms the other methods. This is mostly due to the fact that it can successfully exploit the structure of the covariates. The PLS based variant of the algorithm gives much better results than the principal-component based variant. 

\begin{table}
\centering
\caption{Average $L_2$ error (and standard deviation of the error) obtained for the galaxy example for the different methods compared.}
\label{galaxy_table}       
\begin{tabular}{lrrrr}
\hline\noalign{\smallskip}
  & \multicolumn{2}{c}{Training set} &  \multicolumn{2}{c}{Test set} \\ \hline
%  & \multicolumn{2}{c}{$L_2$ error} &  \multicolumn{2}{c}{L_2$ error} \\
 Using PC directions &$ 1642.00 $&($ 578.4 $)&$ 1758.44 $&($ 615.5 $)\\
 Using PLS directions &$ 511.42 $&($ 79.3 $)&$ 577.05 $&($ 101.9 $)\\
 Linear model & $7420.45$ & ($352.9 $) & $7591.19$ & ($377.5 $) \\
 MARS &$ 2965.64 $&($ 334.7 $)&$ 3738.76 $&($ 494.7 $)\\
 GAM &$ 3027.14 $&($ 321.6 $)&$ 3554.26 $&($ 385.5 $)\\
 PPR &$ 2207.73 $&($ 622.0 $)&$ 3317.94 $&($ 820.7 $)
\\\noalign{\smallskip}\hline
\end{tabular}
\end{table}

%% file: sec5.tex
\section{Application to the Gaia survey mission} 

Gaia is an astrophysics mission of the European Space Agency (ESA) which will
undertake a detailed survey of over $10^9$ stars in our Galaxy and
extragalactic objects.  An important part of the scientific analysis of these
data is the classification of all the objects as well as the estimation of
stellar astrophysical parameters. Below we give a brief overview of the
mission and the data before describing the results obtained using the techniques presented in the preceding section.

\subsection{The astrophysical data}

After launch in 2011, Gaia will study the composition and origin of our Galaxy
by determining the properties of stars in different populations across our
entire Galaxy \cite{ESA:2001,Perryman:01}.  One of its major
contributions will be to measure stellar distances to much higher accuracy
than has hitherto been possible (and will do it for a vast numbers of stars).
Gaia will also measure the three-dimensional space motions of stars in
exquisite detail. These will be used together in dynamical models to map out
the distribution of matter, and can be used to answer fundamental questions
concerning galaxy formation.\footnote{For more information see
  \tt{http://www.rssd.esa.int/Gaia}}

Much of the Gaia astrometric (3D position, 3D velocity) data would
be of little value if we did not know the intrinsic properties of the stars
observed, quantities such as the temperature, mass, chemical compositions,
radius etc.\ (collectively referred to as {\it Astrophysical Parameters}, or
APs; see \cite{Bailer:02b}). For this reason, Gaia is equipped with a low
resolution spectrograph to sample the spectral energy distribution at 96
points across the optical and near-infrared wavelength range (330--1000\,nm).
The measurements themselves are photon counts (energy flux). Each object can
therefore be represented as a point in a 96-dimensional data space. 
 For those objects which are stars, the astrophysical parameters of most 
interest are the following four: (1) {\em effective temperature}, which 
roughly corresponds to the temperature of the observable part of the 
stellar atmosphere; (2) the {\em surface gravity}, which is the 
acceleration due to gravity at the surface of the star; (3) the {\em 
metallicity} or {\em abundance}, a single measurement of the chemical 
composition of the star relative to that of the Sun; (4) the interstellar 
extinction, which measures how much of the star's light has been absorbed 
or scattered by interstellar dust lying between us and the star. In 
practice there is additional ``cosmic variance'' due to other APs, but 
these four are the main ones of interest. (In the rest of this paper we 
examine a simpler case in which there is no variance due to interstellar 
extinction.)

After a century of progress in astrophysics, we now have sophisticated models
of stellar structure and from these we can generate synthetic spectra which
reproduce real stellar spectra reasonably well. Therefore, we can construct
libraries of template spectra with known APs and use these to train supervised
regression models in order to estimate stellar APs. This has received quite a
lot of attention in the astronomical literature, with methods based on nearest
neighbors (e.g. \cite{Prieto:03}), neural networks (e.g. \cite{Baileretal:98},
\cite{Will:05}) and support vector machines (e.g.  \cite{Tsaletal:07})

In its simplest form, the problem of estimating APs with Gaia is one of
finding the optimal mapping between the 96-dimensional data space and the 3 or
more dimensional AP space. In theory this can be solved directly with
regularized nonlinear regression, with the mapping inferred from simulated
data. In practice, however, it is more complicated. For example, the mapping
is not guaranteed to be unique, so some kind of partitioning of the data space
may be appropriate. Furthermore, the intrinsic dimensionality of the data
space is much lower than 96, so we could probably benefit from dimensionality
reduction. Standard PCA has been applied to such data (e.g.\ 
\cite{Baileretal:98}, \cite{Refeetal:07}). It produces more robust models, but
possibly at the cost of filtering out low amplitude features which are
nonetheless relevant for determining the ``weaker'' APs (see below). Here we
investigate local reduction techniques as an alternative.

\subsection{Principal-manifold based approach} 

In this section we study how the methods discussed in Sections 2 and 3 can be
applied to Gaia spectral data. The data comprise several thousand spectra showing variance
in the three astrophysical parameters temperature (in Kelvin), metallicity and
gravity; the latter two variables are on a logarithmic scale.\footnote{In the present example the spectra actually have a
  dimensionality of just 16, rather than 96 as mentioned above, because when
  we carried out this work the Gaia instruments were still being developed.
  Nonetheless, the results we present are illustrative of problems typical in
  observational astronomy.}  Temperature is a ``strong'' parameter, meaning it
accounts for most of the variance across the data set.  Gravity and
metallicity, in contrast, are weak. The parameters have a correlated impact on
the data, e.g.\ at high temperatures, varying the metallicity has a much
smaller impact on spectra than it does at low temperatures.  The data used here are simulated with no noise added.

\begin{figure}
\vspace*{-1cm}
\centerline{\includegraphics[height=10.5cm]{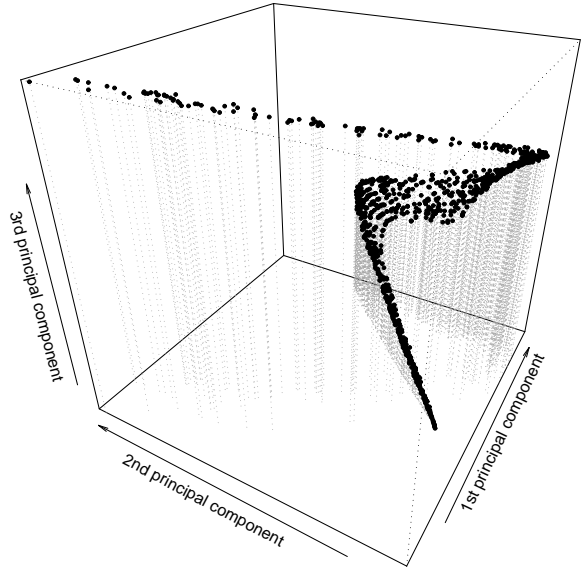}}
\vspace*{-1cm}
\caption{First three principal components of the photon counts\label{photon_counts}}
\end{figure}

The plot of the first three principal components of the covariates, the photon
counts, shows clearly that these possess some low-dimensional structure
(figure \ref{photon_counts}), which cannot be \emph{linearly} approximated.
This suggests employing principal-manifolds based methods.

The low-dimensional structure of the photon counts can be exploited by the projection-based regression tree algorithm (algorithm 5). Recall that the algorithm is based on the idea that the response is mainly determined by the projection of the covariates onto the (tangent to the) principal manifold. This however does not need to be the case; it might well be that the \emph{relevant} information is not captured by the principal manifold.

In the following we will compare whether exploiting the manifold structure of the data allows for obtaining better predictions of the APs. Given the highly nonlinear structure of the data we use support vector regression machines (with a Gaussian kernel) in each partition.\footnote{The optimal cost parameter and the optimal kernel width are determined for each partition individually using a calibration set of 1,000 observations.} The partitions are determined by the PLS-based version of algorithm 5 using 6-dimensional (gravity, metallicity) and 4-dimensional (temperature) hyperplane segments. The results obtained are compared to two further support vector regression machines: one using the original data space of 16 photon counts directly, and one using the first 6 (gravity, metallicity) / 4 (temperature) global principal components as covariates. 

For this purpose, a training set of 3,000 observations, a calibration set of 1,000 observations, and a test set of 1,000 observations is repeatedly drawn from the simulated Gaia data (100 replications). Table \ref{gaia_table} gives the results. As support vector regression machines minimize the $L_1$ distance, we use the $L_1$ error $\sum_{i} |y_i-\hat y_i| / n$. The relative error reported is $\sum_{i} |y_i-\hat y_i| / \sum_{i} |y_i-\bar  y|$.

\begin{table}
\centering
\caption{Average $L_1$ error and relative error (and standard deviation of the error) obtained for the Gaia data.}
\label{gaia_table}       
\subtable[Results obtained for the temperature]{
\begin{tabular}{lrrrrrrrr}
\hline\noalign{\smallskip}
  & \multicolumn{4}{c}{Training set} &  \multicolumn{4}{c}{Test set} \\ 
  & \multicolumn{2}{c}{$L_1$ error} & \multicolumn{2}{c}{Relative err.}  & \multicolumn{2}{c}{$L_1$ error} & \multicolumn{2}{c}{Relative err.}\\  \hline
 Algorithm 5 &  $252.9$ & ($75.1$) & $0.042$ & ($0.012$) & $258.9$ & ($75.1$) & $0.043$ & ($0.013$) \\
 SVR (all counts) & $408.6$ & ($11.6$) & $0.068$ & ($0.002$) & $432.1$ & ($16.0$) & $0.074$ & ($0.004$) \\
 SVR (P.Comps.) & $404.3$ & ($19.6$) &  $0.067$ & ($0.003$) & $412.8$ & ($21.8$) &  $0.070$ & ($0.004$) \\\noalign{\smallskip}\hline
\end{tabular}
}

\vspace*{0.5cm}

\subtable[Results obtained for the gravity]{
\begin{tabular}{lrrrrrrrr}
\hline\noalign{\smallskip}
  & \multicolumn{4}{c}{Training set} &  \multicolumn{4}{c}{Test set} \\ 
  & \multicolumn{2}{c}{$L_1$ error} & \multicolumn{2}{c}{Relative err.}  & \multicolumn{2}{c}{$L_1$ error} & \multicolumn{2}{c}{Relative err.}\\  \hline
%  & $L_1$ error & (sd) &  Rel. error & (sd) & $L_1$ error & (sd) &  Rel. error & (sd) \\ \hline 
 Algorithm 5 &  $0.083$ & ($0.003$) &  $0.069$ & ($0.002$) & $0.109$ & ($0.008$) &  $0.090$ & ($0.005$) \\
 SVR (all counts) & $0.080$ & ($0.001$) & $0.067$ & ($0.001$) & $0.104$ & ($0.005$) & $0.087$ & ($0.005$)\\
 SVR (P.Comps.) & $0.091$ & ($0.002$) & $0.076$ & ($0.001$) &  $0.146$ & ($0.009$) &  $0.118$ & ($0.007$)
\\\noalign{\smallskip}\hline
\end{tabular}
}

\vspace*{0.5cm}

\subtable[Results obtained for the metallicity]{
\begin{tabular}{lrrrrrrrr}
\hline\noalign{\smallskip}
  & \multicolumn{4}{c}{Training set} &  \multicolumn{4}{c}{Test set} \\ 
  & \multicolumn{2}{c}{$L_1$ error} & \multicolumn{2}{c}{Relative err.}  & \multicolumn{2}{c}{$L_1$ error} & \multicolumn{2}{c}{Relative err.}\\  \hline
%  & $L_1$ err. & (sd) & \footnotesize Rel. err. & (sd) & $L_1$ err. & (sd) &  Rel. err. & (sd) \\ \hline 
 Algorithm 5 &  $0.193$& ($0.018$)& $0.134$& ($0.012$)& $0.269$& ($0.016$) & $0.189$ &($0.011$)\\
 SVR (all counts) & $0.279$ & ($0.005$) & $0.193$ & ($0.003$) & $0.363$ & ($0.013$) & $0.253$ & ($0.008$) \\
 SVR (P.Comps.) & $0.256$ & ($0.007$) &  $0.177$ & ($0.004$) & $0.389$ & ($0.014$) & $0.269$ & ($0.011$) \\\noalign{\smallskip}\hline
\end{tabular}
}
\end{table}

The results obtained for the temperature and the metallicity show that using the manifold structure of the photon counts allows for a significant improvement of the predictive performance. The hyperplanes extracted in algorithm 5 seem to capture the information that is relevant to predicting the temperature and the metallicity whilst some of the noise is discarded in the projection step, which facilitates the prediction. This explains why algorithm~5 outperforms the support vector regression machine using all photon counts. The principal components are less able to capture the relevant information; the performance of the support vector regression machine using the first 6 (or 4, respectively) global principal components is clearly worse. 

The results obtained for the gravity, the weakest of the APs,  however give a different picture. Using the manifold structure does \emph{not} allow for improved predictions of the gravity. The information relevant for predicting the gravity seems to be ``orthogonal'' to the extracted hyperplane segments. Thus the support vector regression machine using all the photon counts performs better than the ones based on lower-dimensional projections.

%% file: sec6.tex
\section{Conclusion}

We have reviewed several approaches and algorithms for the representation of high-dimensional complex data structures through lower-dimensional curves or manifolds, which share the property of being based --- by some means or other --- on carrying out principal component analysis ``locally''.  We have focused on localized versions of principal curves (algorithm 4), and on an ``intelligent'' partitioning  algorithm (algorithm 2) avoiding the necessity to specify  an initial partitioning as with usual cluster-wise (``local'') PCA.

We have demonstrated, using the latter of the two algorithms, how localized principal components can be used to reduce the dimension of the predictor space in a non-linear high-dimensional regression problem. The information relevant to predicting the response variable can be elegantly taken into account by maximizing the covariance between the response variable and the extracted projection directions akin to partial least squares (PLS).  We have applied this technique successfully to photon count data from the Gaia mission, where we were able to improve the predictive performance for some of the response variables significantly. The framework presented here however does not guarantee an improved prediction,  especially if the information relevant to the prediction problem at hand cannot be captured by the extracted low-dimensional structure --- as it was the case for the gravity.

It would be desirable to be able to also apply local principal curves (algorithm 4) to such problems. In contrast to algorithm 2, LPCs have the advantage  of representing the covariate space through a proper curve (or manifold) instead of disconnected line (or hyperplane) segments. Currently, algorithm 4 can only extract one-dimensional curves, which it approximates by a sequence of points. Generalizing this idea to higher dimensions, one can approximate a $d$-dimensional principal manifold by a $d$-dimensional mesh, very much like the elastic net algorithm proposed by Gorban and Zinovyev \cite{Gorban:05}. The (basic) elastic net algorithm however is a ``top-down'' method that iteratively bends a mesh of points, starting with a given topology. The generalization of algorithm 4 would use a ``bottom-up'' approach, i.e. learn the local topology from the data requiring no initialization.

However, generalizing algorithm 4 to $d$-dimensional manifolds poses a number of challenges. Firstly, the angle penalization  needs to be modified that it can be applied to local principal components of higher order. Further one has to make sure that the different branches meet, forming a proper mesh of points, which will require keeping the distance between subsequent $\vec{\mu}^{\vec{x}}$ constant.